\documentclass[twocolumn]{aastex631}
\usepackage{natbib}
\bibpunct{(}{)}{;}{a}{}{,} 
\usepackage{amssymb,amsmath}
\usepackage{mathtools}
\usepackage{soul}
\usepackage{txfonts}
\usepackage{graphicx}
\usepackage{color}
\definecolor{darkgreen}{rgb}{0,0.60,0}

\DeclarePairedDelimiterX\braket[2]{\langle}{\rangle}{#1 \delimsize\vert #2}
\usepackage[nonumberlist,nosuper]{glossaries}
\setacronymstyle{long-short}
\newacronym{1d}{1D}{one-dimensional}
\newacronym{3d}{3D}{three-dimensional}
\newacronym{clv}{CLV}{center-to-limb variation}
\newacronym{fwhm}{FWHM}{full width at half maximum}
\newacronym{fov}{FOV}{field of view}
\newacronym{fs}{FS}{fine structure}
\newacronym{hfs}{HFS}{hyperfine structure}
\newacronym{los}{LOS}{lines of sight}
\newacronym{lte}{LTE}{local thermodynamical equilibrium}
\newacronym{mo}{MO}{magneto-optical}
\newacronym{prd}{PRD}{partial frequency redistribution}
\newacronym{rt}{RT}{radiative transfer}
\newacronym{rte}{RTE}{radiative transfer equations}
\newacronym{see}{SEE}{statistical equilibrium equations}
\newacronym{sj}{SJ}{slit-jaw}
\newacronym{tr}{TR}{transition region}
\newacronym{uv}{UV}{ultraviolet}
\newacronym{wfa}{WFA}{weak field approximation}
\shorttitle{Potential of filter-polarimetry in Ly-alpha}
\shortauthors{Alsina Ballester et al.}
\submitjournal{ApJ}
\begin{document} 
\title{The potential of the wavelength-integrated scattering polarization of \\the hydrogen Ly$\alpha$ 
line for probing the solar chromosphere}
\correspondingauthor{E. Alsina Ballester}
\email{ernest.alsina@iac.es}
\author[0000-0001-9095-9685]{E. Alsina Ballester}
\affiliation{Instituto de Astrof\'{i}sica de Canarias (IAC), 
E-38205 La Laguna, Tenerife, Spain}
\affiliation{Departamento de Astrof\'{i}sica, 
 Universidad de La Laguna {(ULL)}, 
 E-38206 La Laguna, Tenerife, Spain}
\author[0000-0002-8775-0132]{L. Belluzzi}
\affiliation{Istituto ricerche solari Aldo e Cele Dacc\`o (IRSOL), 
Universit\`a della Svizzera italiana (USI), 
CH-6605 Locarno-Monti, Switzerland}
\affiliation{Leibniz-Institut f\"{u}r Sonnenphysik (KIS), 
D-79104 Freiburg, Germany}
\affiliation{Euler Institute, Universit\`a della Svizzera italiana (USI),
CH-6900 Lugano, Switzerland}
\author[0000-0001-5131-4139]{J. Trujillo Bueno}
\affiliation{Instituto de Astrof\'{i}sica de Canarias (IAC), 
E-38205 La Laguna, Tenerife, Spain}
\affiliation{Departamento de Astrof\'{i}sica, 
 Universidad de La Laguna {(ULL)}, 
 E-38206 La Laguna, Tenerife, Spain}
 \affiliation{Consejo Superior de Investigaciones Cient\'{i}ficas (CSIC), 
  Spain}
\begin{abstract}
{The intensity and the linear scattering polarization profiles of the hydrogen Ly$\alpha$ line encode valuable information on the thermodynamic and magnetic structure of the upper layers of the solar chromosphere. 
The Chromospheric Lyman-Alpha Spectro-Polarimeter (CLASP) sounding rocket experiment provided unprecedented spectropolarimetric data of this line, as well as two-dimensional broadband images in intensity and linear polarization.} 
We theoretically investigate the potential of the Ly$\alpha$ broadband polarimetric signals for probing the solar chromosphere and its magnetic fields. 
{We analyze the synthetic Stokes profiles obtained from a series of radiative transfer (RT) calculations out of local thermodynamic equilibrium, considering semi-empirical one-dimensional models of the solar atmosphere.} 
{The wavelength-integrated linear polarization signal is found to be dominated by the contribution from the wings when considering a Gaussian weighting function with a 
FWHM that corresponds to the CLASP slit-jaw broadband filter. These broadband linear polarization signals are strongly sensitive to magnetic fields of strengths on the order of $50$~G, via the action of magneto-optical (MO) effects, and are expected to encode information on the middle-upper chromosphere.} 
The two-dimensional broadband intensity and linear polarization images observed by CLASP can be suitably mimicked using synthetic wavelength-integrated signals obtained considering atmospheric models and magnetic fields that are representative of solar regions with different levels of activity, provided that the impact of MO effects is taken into account. Despite the limitations of a one-dimensional RT modeling, this work illustrates the diagnostic potential of filter-polarimetric Ly$\alpha$ signals for probing the solar chromosphere and its magnetism. 
\end{abstract}
 \date{\today}
 \keywords{Scattering polarization --
                spectral lines --
                radiative transfer -- 
                Sun: chromosphere }
\section{Introduction} 
In the solar chromosphere-corona \gls*{tr}, the temperature suddenly rises from $\sim~\!\!\!\!10^4$\,K to $\sim~\!\!\!\!10^6$\,K. During recent years, a number of physical explanations for this phenomenon have been proposed, highlighting the need for more detailed observational data on the thermal and magnetic properties of the TR and underlying layers. 
The hydrogen Ly$\alpha$ resonance line (also referred to as Lyman-$\alpha$ in the literature), the strongest emission line in the solar ultraviolet spectrum, represents an especially valuable observational window in this regard. The near-wing and line-center photons of this line encode information on the upper chromosphere and the regions immediately below the \gls*{tr}.  
Radiative transfer (RT) calculations showed that scattering processes produce measurable linear polarization signals in the hydrogen Ly$\alpha$ line \citep{TrujilloBueno+11,Belluzzi+12,Stepan+15}. 
Because the scattering polarization depends on the anisotropy of the radiation field \citep[e.g.,][]{TrujilloBueno01}, it encodes information on the thermal and geometrical structure of the atmospheric regions from which the observed radiation is emitted.  
Moreover, the scattering polarization signals found in the line core are sensitive to the Hanle effect, thereby providing a window into the elusive magnetic fields of the upper chromospheric layers, just below the \gls*{tr}. These theoretical predictions motivated the development of the Chromospheric Lyman-Alpha Spectro-Polarimeter (CLASP) sounding rocket experiment, which carried out observations of the intensity and linear polarization of the solar H~{\sc{i}} Ly$\alpha$ line with the slit oriented radially, spanning positions from close to the disk center to the limb \citep[see][]{Kano+17}. 
Exploiting the unprecedented spectropolarimetric data provided by CLASP, it was possible to set constraints on the magnetization and the degree of corrugation of the \gls*{tr} \citep{TrujilloBueno+18}. 

The radiation pertaining to the core and near wings of Ly$\alpha$ originates from just below the \gls*{tr} and from the upper chromosphere, whereas the radiation farther into the wings originates from deeper layers, but still at chromospheric heights \cite[see Fig.~1 of][]{Belluzzi+12}. For instance, in the semi-empirical model C of \cite{Fontenla+93}, the radiation at $4$~\AA\ from the line core is found to come primarily from heights around $1250$~km above the photospheric surface. 
Large-amplitude scattering polarization signals, produced by partially coherent scattering processes, are found in the line wings. 
These linear polarization signals can be suitably modeled by accounting for the frequency correlations between incident and scattered radiation (i.e., \gls*{prd} effects) and the quantum interference between the \gls*{fs} levels upper term of the line transition (i.e., $J$-state interference). 
Although these wing scattering polarization signals are largely insensitive to both the Hanle and the familiar Zeeman effect (see Sect.~10.4 of \citealt{BLandiLandolfi04}, also \citealt{AlsinaBallester+19}), they are nevertheless impacted by magnetic fields with a significant longitudinal component through the action of \gls*{mo} effects. 
Specifically, such wing signals are modified by the \gls*{mo} effects described by the $\rho_V$ coefficient of the \gls*{rt} equation, which induces a rotation of the plane of linear polarization \citep[see][]{AlsinaBallester+16} and can also give rise to an effective decrease in its linear polarization fraction \citep[see Appendix A of][]{AlsinaBallester+18}. 
Within the line-core region, it is well known that the impact of such MO effects is only significant in the presence of magnetic fields with strengths comparable to those for which the impact of the Zeeman effect is appreciable in the linear polarization \citep[e.g., Sect.~9.22 of][]{BLandiLandolfi04}. 
On the other hand, a few years ago it was shown that, outside the line core, the $\rho_V$ coefficient becomes comparable to the absorption coefficient $\eta_I$ in the presence of longitudinal magnetic fields whose strength is similar to the one characterizing the onset of the Hanle effect \citep[see Appendices B and C of][]{AlsinaBallester+19}. Thus, the wing scattering polarization signals are expected to be sensitive to relatively weak solar magnetic fields. For the Ly$\alpha$ case, a clear depolarization of $Q/I$ and the appearance of a $U/I$ signal was predicted for magnetic fields substantially weaker than $50$~G \citep{AlsinaBallester+19}\footnote{That RT investigation was carried out considering a spinless two-level atomic model, after numerically demonstrating its suitability. The minor differences found in the near wings of the $\rho_V/\eta_I$ ratio, relative to the two-term calculation, were simply an artifact due to improper normalization in the latter case. With a correct normalization, the ratios found for the two calculations are indistinguishable  throughout the considered spectral range.}. 

\begin{figure}[!h]
 \centering
\includegraphics[width=8.25cm]{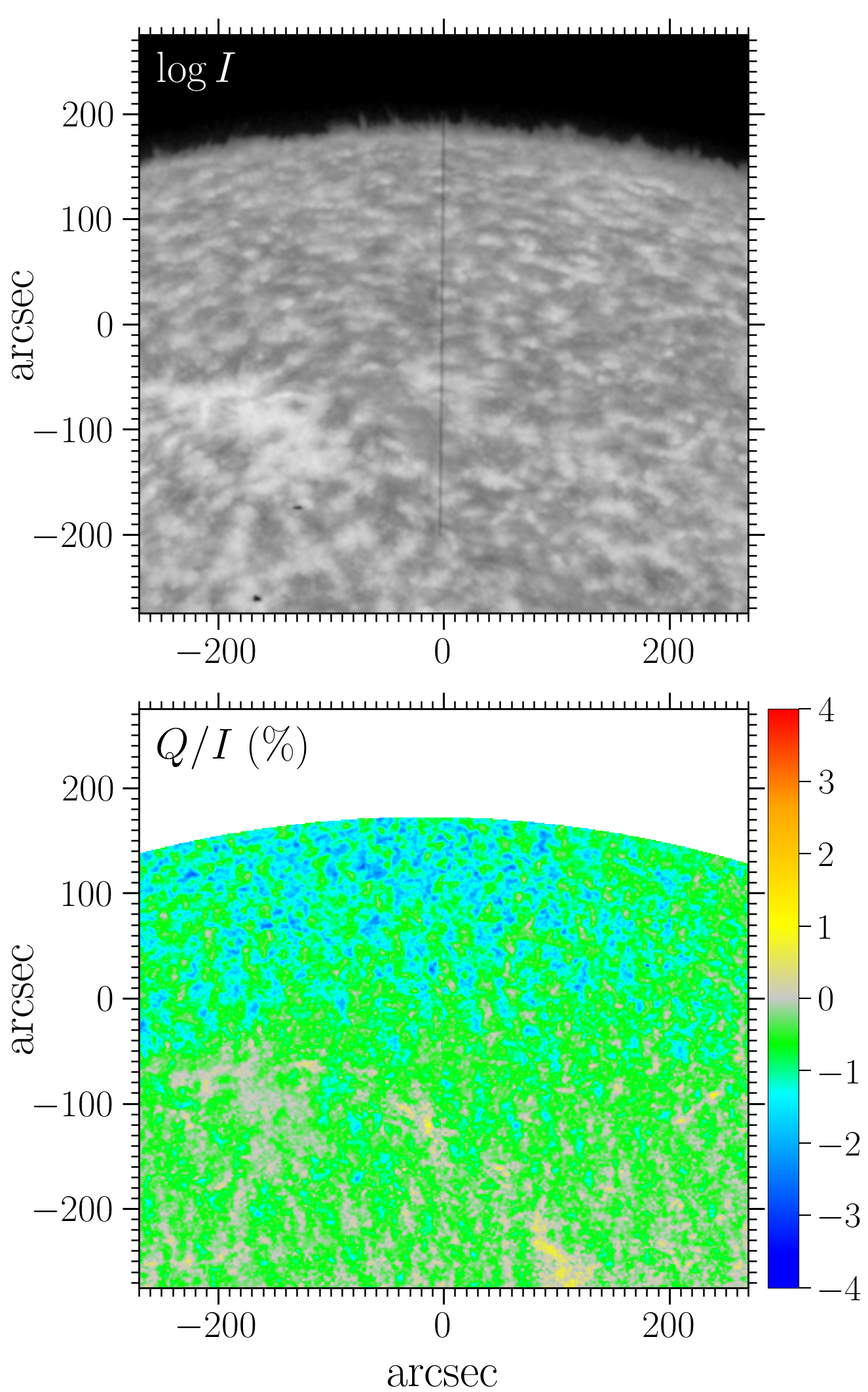}
\caption{Stokes $I$ (\textit{upper panel}) and $Q/I$ (\textit{lower panel}) images taken by the CLASP slit-jaw (SJ) camera, whose broadband filter was centered on the H~{\sc {i}} Ly$\alpha$ line, with a \gls*{fwhm} of about $35$~\AA . The reference direction for positive Stokes $Q$ was taken parallel to the spectrograph slit, which is visible in the upper panel. Figure adapted from \cite{Kano+17}, after correcting a small error in the color table of the lower panel.} 
\label{fig::CLASP_SJImage}
\end{figure}
In addition to the above-mentioned spectropolarimetric data, CLASP also performed broadband polarimetric measurements over a large \gls*{fov} with its \gls*{sj} system. {The \gls*{sj} observation was centered on a region of the solar disk close to the solar limb that contained a bright plage area and a multitude of network features. Serendipitously, the \gls*{sj} provided broadband images in $Q/I$ as well as in intensity, as shown in Fig.~\ref{fig::CLASP_SJImage}. These measurements revealed that the plage and network features seen in the intensity image (which are generally associated with higher magnetic activity) show much lower broadband linear polarization amplitudes than the surrounding areas. 

The main aim of this work is to investigate the diagnostic potential of the broadband intensity and linear polarization signals centered on the Ly$\alpha$ line, placing special emphasis on the magnetic sensitivity of the linear polarization through the action of the Hanle, Zeeman, and \gls*{mo} effects. 
In recent years, investigations carried out with the \gls*{3d} \gls*{rt} code PORTA \citep{StepanTrujilloBueno13} highlighted the crucial impact on the scattering polarization  
of phenomena that are inherent to the complex \gls*{3d} and dynamic nature of the solar atmosphere, including 
horizontal inhomogeneities in atmospheric properties such as temperature or density and spatial gradients in the 
horizontal component of bulk velocities \citep[e.g.,][]{delPinoAleman+18,JaumeBestard+21}. 
Unfortunately, no \gls*{3d} \gls*{rt} code currently exists that can account for scattering polarization with \gls*{prd} effects and $J$-state interference. Therefore, the present work can be considered to be a first step towards the goal of investigating the combined action of the MO effects (whose impact is analyzed in the present paper) and the symmetry breaking caused by the horizontal inhomogeneities of the solar chromosphere (which requires a 3D RT code).  

In Sect.~\ref{sec::Formulation}, we briefly discuss the atomic model and the numerical \gls*{rt} code used to synthesize the intensity and polarization profiles of the Ly$\alpha$ line. In Sect.~\ref{sec::Results}, we analyze the synthetic intensity and linear polarization patterns of Ly$\alpha$ that are obtained with different semi-empirical atmospheric models, whose thermodynamic properties are representative of different solar regions, and discuss their magnetic sensitivity. 
In the same section, we integrate the profiles over wavelength, weighting them by a Gaussian function that corresponds to the CLASP \gls*{sj} broadband filter, and analyze the contribution from different spectral regions to the integrated intensity and polarization signals. 
We conclude Sect.~\ref{sec::Results} by highlighting the diagnostic potential of the Ly$\alpha$ broadband signals. 
For illustrative purposes, we use the wavelength-integrated synthetic Stokes profiles to mimic the large-\gls*{fov} broadband images obtained with CLASP. 
Section~\ref{sec::Conclusions} lays out the conclusions of this work and future perspectives. 

\section{Formulation of the problem}
\label{sec::Formulation} 
The hydrogen Ly$\alpha$ line arises from radiative transitions between states with principal quantum numbers $n=2$ and $n=1$. 
The states with $n=1$ can only have orbital quantum number $l=1$ ($1s$), whereas those with $n=2$ can have $l=1$ ($2s$) and $l=2$ ($2p$). 
Taking the \gls*{fs} of hydrogen into account, the states in the $2p$ configuration belong to two different \gls*{fs} levels, with total angular momentum $j = 1/2$ and $j = 3/2$, whereas the states in the $1s$ and $2s$ configurations each belong to a single \gls*{fs} level with $j = 1/2$. 
The transitions between $2s$ and $1s$ states are prohibited by the selection rules of electric-dipole type interactions and can be safely neglected. 
We neglect the \gls*{hfs} of hydrogen in this work, bearing in mind that the frequency separation between the \gls*{hfs} levels of the considered \gls*{fs} levels is much smaller than their natural width (\citealt{BommierSahalBrechot82}; see also Sect.~3.1.5 of \citealt{TrujilloBueno+17}).  
In this work, we thus model the Ly$\alpha$ line considering a two-term (${}^2S \, - \, {}^2P$) model atom for hydrogen. 
This model allows taking into account the quantum interference between the ${}^2 P_{1/2}$ and ${}^2 P_{3/2}$ upper \gls*{fs} levels, which were shown to have a substantial impact on the scattering polarization wings of Ly$\alpha$ \citep[see][]{Belluzzi+12}.
Because the ${}^2 S_{1/2}$ lower level is the ground level of hydrogen, which has a very long lifetime, we can safely assume that it infinitely sharp and unpolarized. 

The intensity and polarization patterns of the solar Ly$\alpha$ line presented in the following sections are synthesized by numerically solving the RT problem for polarized radiation, out of \gls*{lte} conditions, in semi-empirical \gls*{1d} models of the solar atmosphere. 
The calculations are carried out with the code introduced in \cite{AlsinaBallester+22}, hereafter ABT22. The code is based on the theoretical framework presented in \citet{Bommier17} and can account for scattering polarization and for \gls*{prd} phenomena (both collisional and Doppler) in a two-term atomic system. The code can also consider magnetic fields of arbitrary strength and orientation in the incomplete Paschen-Back effect regime, thus accounting for the combined action of the Hanle, Zeeman, and \gls*{mo} effects.

The emergent radiation is calculated by solving the \gls*{rte} for polarized radiation \citep[e.g.,][]{BLandiLandolfi04} 
\begin{equation}
    \frac{\mathrm{d}}{\mathrm{d}s} \boldsymbol{I} = \boldsymbol{\varepsilon} - \mathbf{K} \, \boldsymbol{I} \, , 
\label{Eq:RT}
\end{equation}
where $\boldsymbol{I}=(I,Q,U,V)^T$ is the Stokes vector and $s$ the spatial coordinate along the ray path. 
Assuming stationary conditions, the quantities appearing in Eq.~\eqref{Eq:RT} depend in general on the spatial point $\mathbf{r}$ and on the frequency $\nu$ and the propagation direction $\mathbf{\Omega}$ of the considered radiation beam. These dependencies are not given explicitly for simplicity of notation. 
The propagation matrix $\mathbf{K}$ is a $4 \times 4$ matrix of the form
\begin{equation}
    \mathbf{K} = 
    \left(
    \begin{array}{c c c c}
        \eta_I & \eta_Q & \eta_U  & \eta_V \\
        \eta_Q & \eta_I & \rho_V  & -\rho_U \\
        \eta_U & -\rho_V & \eta_I & \rho_Q \\
        \eta_V & \rho_U & -\rho_Q & \eta_I
    \end{array} \right) \, . 
\label{Eq:PropMat}
\end{equation}
The diagonal element $\eta_I$ is the absorption coefficient for intensity, the elements $\eta_X$ with $X = \{Q, U, V\}$ describe the differential absorption of the corresponding Stokes parameter (i.e., dichroism), and the anomalous dispersion coefficients $\rho_X$ characterize the couplings between the Stokes components other than $I$. The elements of the propagation matrix and the emission vector are collectively referred to as the \gls*{rt} coefficients. The code takes into account both the line and continuum contributions to each of them; the continuum contributions are discussed in detail in ABT22. 

The line contribution to the emission vector is itself given by the sum of two terms that describe the thermal and scattering contributions, respectively. 
The explicit form of the thermal term is presented and discussed in \cite{AlsinaBallester22}. 
The scattering term is calculated using the redistribution matrix formalism:
\begin{equation}
    \boldsymbol{\varepsilon}^{\ell, {\rm sc}}(\nu,\mathbf{\Omega}) = k_L \int \! \mathrm{d} \nu^\prime \oint \! \frac{{\mathrm d} \mathbf{\Omega}'}{4 \pi} 
    \boldsymbol{\mathcal{R}}(\nu',\mathbf{\Omega}',\nu,\mathbf{\Omega}) \, \boldsymbol{I}(\nu',\mathbf{\Omega}') \, ,
\label{Eq:eps_sc}
\end{equation} 
where the convention according to which primed quantities refer to the incident radiation and unprimed ones to the emitted radiation is used. The quantity $k_L$ is the frequency-integrated absorption coefficient and $\boldsymbol{\mathcal{R}}$ is the redistribution matrix. We recall that the redistribution matrix formalism is only applicable when the \gls*{see} have an analytical solution \citep[see ABT22, also][]{Bommier18}, which is implicitly contained in $\boldsymbol{\mathcal{R}}$. 
This is indeed the case for the considered two-term atomic system with an infinitely sharp and unpolarized lower term, provided that stimulated emission and the collisions that couple different $J$ levels of the same term are neglected \citep{Bommier17}. The redistribution matrix is given by the sum of two terms, which characterize scattering processes that are coherent ($\boldsymbol{\mathcal{R}}_{\mbox{\sc{ii}}}$) and completely incoherent ($\boldsymbol{\mathcal{R}}_{\mbox{\sc{iii}}}$) in the atomic rest frame. In the observer’s frame, we consider for computational simplicity the so-called angle-averaged approximation for $\boldsymbol{\mathcal{R}}_{\mbox{\sc{ii}}}$ \citep{ReesSaliba82}, which we expect to be suitable for modeling the wings of resonance lines \citep[see][]{Janett+21}. 
We also make the assumption that scattering is completely incoherent in the observer’s frame for $\boldsymbol{\mathcal{R}}_{\mbox{\sc{iii}}}$. 

By applying suitable iterative methods (see ABT22), the code reaches a self-consistent solution of the coupled Eqs.~\eqref{Eq:RT} and \eqref{Eq:eps_sc}. 
Under the assumptions of an unpolarized lower term and of neglecting stimulated emission, $k_L$ and the elements of $\mathbf{K}$ depend on the atomic state only through the population of the lower term $\mathcal{N}_\ell$ \citep[e.g.,][]{BLandiLandolfi04}. In our approach, we keep $\mathcal{N}_\ell$ fixed throughout the iterative calculation, as discussed in detail in ABT22. 
In this work, we use the hydrogen ground level population provided by the considered atmospheric models.
\vfill
\section{Results}
\label{sec::Results} 
In this section we present the results of a series of \gls*{rt} calculations, in order to highlight the diagnostic potential of the broadband Ly$\alpha$ intensity and linear polarization signals for probing the properties of the solar chromosphere, placing particular emphasis on its magnetic fields.  
For such calculations, we consider the \gls*{1d} semi-empirical models presented in \cite{Fontenla+93}, which we will hereafter refer to as FAL models (for instance, model A will be referred to as FAL-A). 
We specify the \gls*{los} of the emergent radiation simply through parameter $\mu = \cos\theta$, where $\theta$ is the inclination with respect to the local vertical which, for the considered geometry, corresponds to the heliocentric angle. 
Because static plane-parallel atmospheric models are considered, the axial symmetry of the problem can only be broken in the presence of an inclined magnetic field. When magnetic fields are included in the \gls*{rt} calculations, the same strength and orientation is taken at all heights in the model. 
The orientation is given by the inclination of the field $\theta_B$, with respect to the local vertical, and by its azimuth $\chi_B$, relative to the plane defined by the \gls*{los} and the local vertical. This geometry is illustrated in Fig.~1 of ABT22. In the RT calculations presented below, the reference direction for positive Stokes $Q$ is taken parallel to the limb unless otherwise noted. 

\subsection{Sensitivity of the Stokes profiles to the thermodynamic and magnetic structure of the solar atmosphere}
\label{subsec::Profiles}
Considerable variations of the thermodynamical and magnetic properties of the solar atmosphere should be expected within the large \gls*{fov} encompassed by the two-dimensional broadband images observed by the CLASP \gls*{sj} instrument. 
In this subsection, we investigate how the synthetic Ly$\alpha$ intensity and linear polarization profiles vary according to the considered atmospheric model in the \gls*{rt} calculations. 
Figure~\ref{fig:FigModel} shows a comparison between the intensity and Stokes $Q/I$ profiles obtained in the absence of magnetic fields considering the following atmospheric models: FAL-A, which is representative of particularly faint internetwork regions of the quiet Sun; 
FAL-F, representative of relatively bright network regions of the quiet Sun; FAL-C, representative of regions of intermediate brightness; and FAL-P, representative of a typical plage region. 
In order to obtain scattering polarization signals with a large amplitude, the profiles are shown for an \gls*{los} that corresponds to a large inclination, namely with $\mu = 0.3$. 
For all of the considered models, we find an increase of roughly two orders of magnitude in the Ly$\alpha$ intensity profile when going from the far line wings into the core region\footnote{We define the core region as the spectral range up to $200$~m\AA\ from the Ly$\alpha$ line center.}. On the other hand, the intensity itself strongly varies depending on the considered model. Indeed, the intensity values obtained with FAL-P are larger than those obtained with FAL-C by roughly one order of magnitude, both in the line core and the line wings. 
\begin{figure}[!h]
 \centering 
\includegraphics[width=8.25cm]{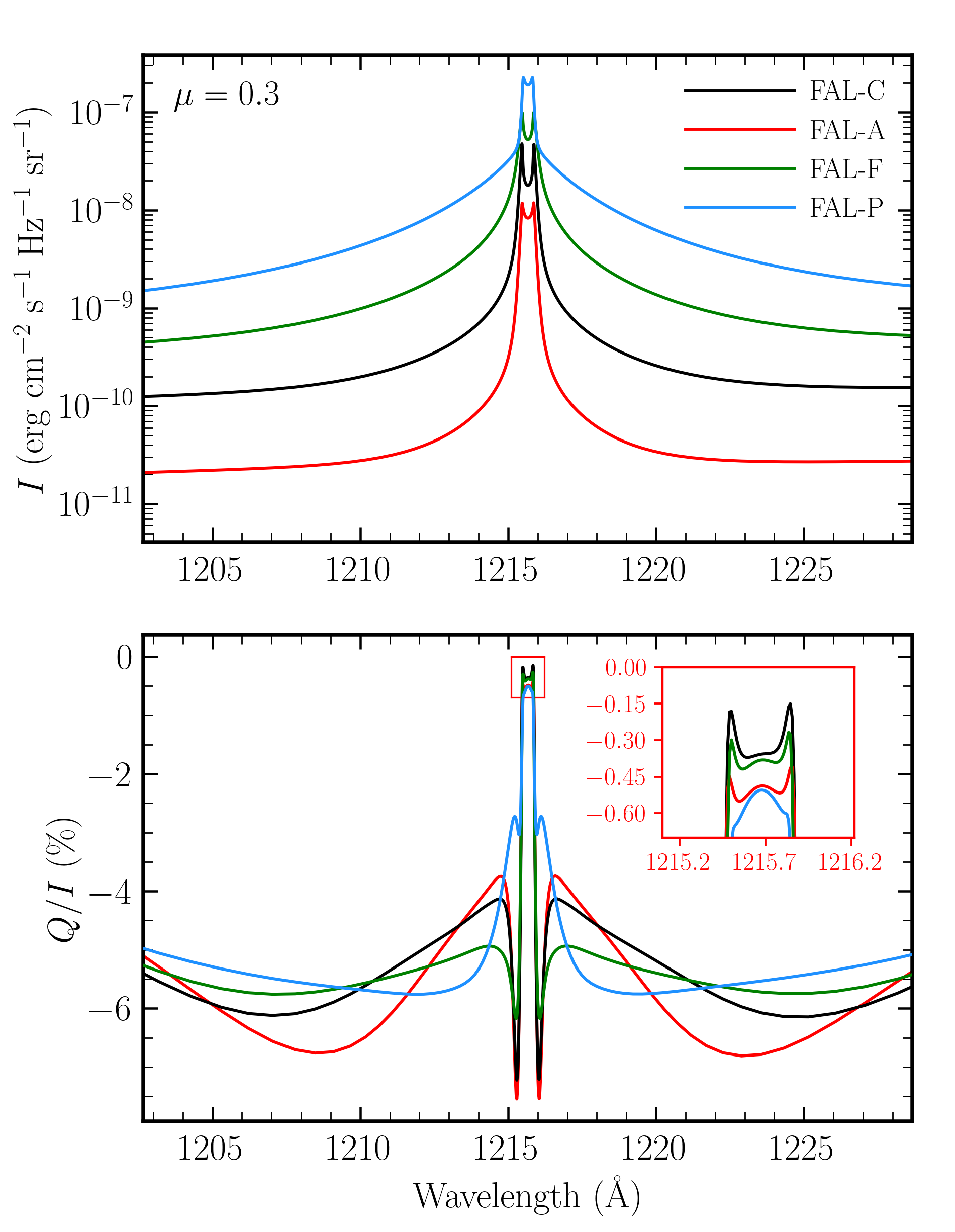}
\caption{Stokes $I$ (\textit{upper panel}) and $Q/I$ (\textit{lower panel}) profiles for the H~{\sc{i}} Ly$\alpha$ line, obtained via the \gls*{rt} calculations described in the text in the absence of magnetic fields, as a function of vacuum wavelength. The colored curves indicate the results of the calculations considering the atmospheric models A, C, F, and P of \cite{Fontenla+93}, as indicated in the legend. The inset figure in the lower panel shows the same $Q/I$ profiles, but in a narrower wavelength interval. An LOS with $\mu = \cos\theta = 0.3$ is considered and the direction for positive Stokes $Q$ is taken parallel to the limb.}  
\label{fig:FigModel}
\end{figure} 

The linear polarization profiles obtained for the various models have qualitative similarities; they present a relatively small amplitude in the core, large-amplitude negative peaks in the near wings, and broad lobes of slightly smaller amplitude farther into the wings. Throughout the considered wavelength range, the profile is negative, as a consequence of a negative radiation anisotropy (i.e., the intensity is stronger for the radiation that propagates horizontally than vertically) in the spatial regions from which both the line core and wing photons originate. 
The largest line-core amplitude is found when considering FAL-P and the smallest is found when considering FAL-C, but for all models it is well below $1\%$ in absolute value. 
The amplitudes of both the near-wing $Q/I$ peaks and the lobes farther from the line core depend strongly on which model is considered. The largest wing amplitudes are found, in descending order, for FAL-A, C, F, and finally P. 
Comparing the profiles obtained with the latter three models to the corresponding profiles in Fig.~4 of \cite{Belluzzi+12}, we find an excellent agreement in the line core and in the negative peaks in the near wings. However, there are appreciable discrepancies between the signals found in the outer lobes, which we attribute to differences in the branching ratios between the formulations used for the two numerical approaches (for details, see ABT22), leading to a slight overestimation of the polarization introduced by $J$-state interference in \cite{Belluzzi+12}. 

We also carried out a series of \gls*{rt} calculations considering the same atmospheric models, but in the presence of magnetic fields with varying strengths on the order of the Hanle critical field of the line \citep[roughly $53$~G, see][]{TrujilloBueno+12}. 
The magnetic fields were taken to be deterministic and uniform throughout the atmospheric model. For the sake of brevity, we do not show in this subsection the resulting Stokes profiles,\footnote{The Stokes profiles calculated in the presence of horizontal magnetic fields in the FAL-C model can be found in \citet{AlsinaBallesterSPW9}.}  but they are used in the following subsections. 
We note here that the intensity patterns are not appreciably impacted by magnetic fields of such strengths, and thus they depend mainly on variations in other thermodynamical properties. 
The $Q/I$ and $U/I$ signals in the core region are sensitive to the presence of inclined magnetic fields through the action of the Hanle effect, which tends to depolarize and rotate the linear polarization signals close to the limb \citep[see Figs.~3 and 4 of][]{TrujilloBueno+11}. 
Outside the line core region, $Q/I$ and $U/I$ are sensitive, through the action of \gls*{mo} effects, to longitudinal magnetic fields with strengths comparable to the Hanle critical field \citep[e.g., Fig.~2 of][]{AlsinaBallester+19}. 
The discussion on the circular polarization signals is omitted from this work, because they present a relatively small amplitude until strengths of a few hundred gauss are considered. Moreover, such signals tend to cancel when integrating over a wavelength interval that is symmetric around the line center, as is done in the following subsections.  

 \subsection{Wavelength-integrated Stokes signals}
 \label{subsec::Freqint} 
 \begin{figure*}[!t]
\centering
\includegraphics[width=16cm]{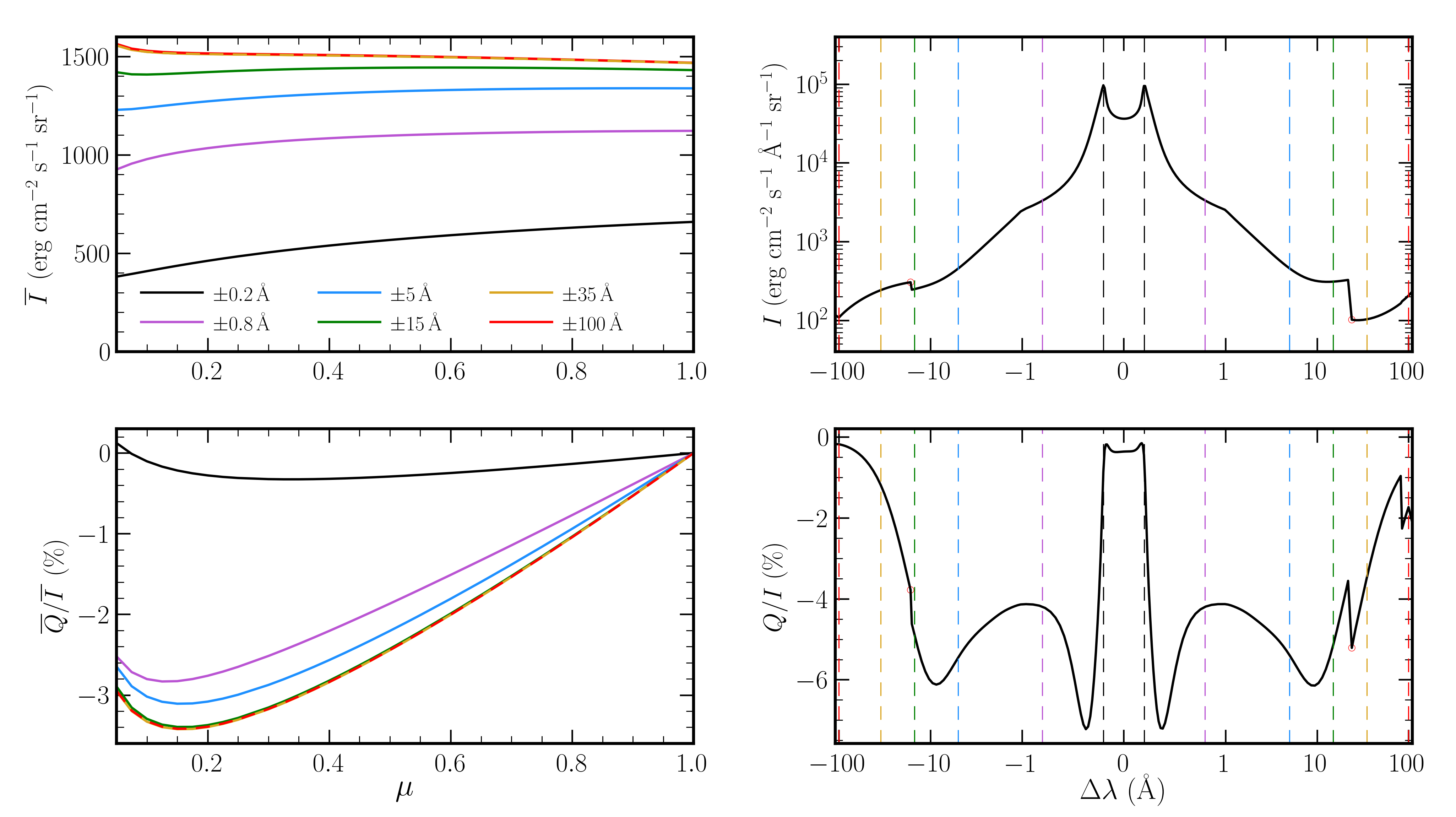} 
\caption{\textit{Left panels}: \gls*{clv} of the broadband $\overline{I\,}$ (\textit{upper left}) and $\overline{Q}/\overline{I\,}$ (\textit{lower left}) signals as a function of $\mu = \cos\theta$. 
These signals are obtained by wavelength-integrating the synthetic Stokes profiles obtained from an \gls*{rt} calculation considering the FAL-C atmospheric model in the absence of magnetic fields, as discussed in the text. 
The various colored curves represent the \gls*{clv}s obtained when considering different limits for the wavelength integration (see legend). Overlapping curves are dashed for the sake of clarity. 
\textit{Right panels}: Intensity (\textit{upper right}) and $Q/I$ (\textit{lower right}) profiles that enter the wavelength integration, given as a function of wavelength distance $\Delta \lambda = \lambda - \lambda_0$, where $\lambda_0 = 1215.67$~\AA\ is the line-center wavelength. For visualization purposes, a linear scale is taken within $|\Delta\lambda| = 1$~\AA\ and a logarithmic one is taken farther from line center. The vertical dashed colored lines indicate the wavelength intervals spanned by each of the aforementioned integration intervals. The direction for positive Stokes $Q$ is taken parallel to the limb.} 
\label{fig:Intprof}
\end{figure*}
In this subsection we focus on the theoretical Ly$\alpha$ intensity and polarization signals  
obtained using broadband filters, such as those installed on the CLASP \gls*{sj} system, which have a \gls*{fwhm} of $35$~\AA . 
From the synthetic Stokes-vector profiles $\mathbf{I}\, (\nu,\mathbf{\Omega})$ discussed in the previous subsection, we can obtain the broadband Stokes vector 
$\mathbf{\overline{I\,}} (\mathbf{\Omega})$ by integrating over wavelength,  
 \begin{equation}
 \overline{I\,}_{\!i}\,(\mathbf{\Omega}) = \frac{1}{\sqrt{2\pi}\sigma_{\!\lambda}} 
\int_{\lambda_0-{\Delta\lambda_L}}^{\lambda_0+{\Delta\lambda_L}}  
 \!\!\!\mathrm{d}\lambda \, \exp\,\Biggl(-\frac{1}{2}\biggl(\frac{\lambda - \lambda_0}{\sigma_{\!\lambda}} \biggr)^2 \Biggr) \; I_i(\lambda,\mathbf{\Omega}) \, , 
 \label{Eq::WaveInt}
 \end{equation}
 where the index $i = \{0,1,2,3\}$ corresponds to the $I$, $Q$, $U$, $V$ components of the Stokes vector, respectively. $\lambda_0 = {1215.67}$~{\AA} is the Ly$\alpha$ line center in vacuum wavelength and the integration limits are taken to be symmetric around $\lambda_0$, so that the full interval is $2\,\Delta\lambda_L$. The standard deviation of the Gaussian weighting function $\sigma_{\!\lambda}$ is selected so that it corresponds to a \gls*{fwhm} of $35$~\AA , noting that $\mathrm{FWHM} = 2\sqrt{2 \ln{2}} \, \sigma_{\!\lambda}$. 
The synthetic $I_i(\lambda,\mathbf{\Omega})$ profiles were obtained on a wavelength grid with $222$ points, such that the $18$ points within a $60$~m\AA\ interval from $\lambda_0$ are approximately equally spaced and, beyond this interval, the separation between wavelength points is constant on a logarithmic scale. 
 In order to carry out the wavelength integration, the synthetic Stokes profiles were first linearly interpolated onto a refined grid. For the refined grid, the same number of intervals was taken as in the original grid in the wavelength range up to $0.2$~{\AA}, but forcing the wavelength points to be exactly equally spaced. 
 Outside this wavelength range, the refined grid was also taken to be uniform, but in a piecewise manner, taking a step size of $10$\,m\AA\ between $0.2$ and $1.5$\,{\AA} and of $50$\,m\AA\ beyond $1.5$\,\AA . 

For illustrative purposes, we first consider the profiles obtained from a \gls*{rt} calculation in the absence of magnetic fields, considering the FAL-C atmospheric model. In order to evaluate the contribution to the broadband intensity $\overline{I\,}$ and to the ratio $\overline{Q}/\overline{I\,}$ from various wavelength ranges, we compare the synthetic broadband signals obtained by taking different integration limits as specified by $\Delta\lambda_L$. 
In particular, we consider intervals that extend $0.2$~\AA , $0.8$~\AA , $2.4$~\AA , $15$~\AA , $70$~\AA , and $100$~\AA\ 
from line center. 
The \gls*{clv} of the resulting signals is shown in the left panels of Fig.~\ref{fig:Intprof} for \gls*{los}s with $\mu$ between $0.1$ and $1$. For reference, the right panels show the synthetic profiles, as a function of wavelength, for an LOS with $\mu = 0.3$. The vertical dashed colored curves indicate the considered wavelength intervals (see legend), which are given in terms of the distance from line center $\Delta\lambda = \lambda - \lambda_0$. A linear scale is taken for $|\Delta\lambda| \le 1$~\AA\ and a logarithmic scale is taken beyond this spectral distance. %
The jumps in intensity found at $\lambda = 1199.57$ \AA\ ($\Delta\lambda = -16.10$ \AA) and $\lambda = 1239.61$ \AA\ ($\Delta\lambda = 23.94$ \AA), indicated with small red circles, coincide with the photoionization thresholds for the ground level of S~{\sc{i}} and the $2\mbox{s}^2 2\mbox{p}^2 \, {}^{1}D_2$ level of C~{\sc{i}}, respectively. 

Even though the H~{\sc{i}} Ly$\alpha$ line presents a very sharp emission profile, the radiation from the core region represents about a third of the total contribution to 
$\overline{I\,}$ (see black curve in the upper left panel of Fig.~\ref{fig:Intprof}). The interval between $200$ and $800$~m\AA\ from the line center contributes roughly another third of the total $\overline{I\,}$. We note that a limb darkening is found when considering these narrow intervals but not for wider spectral ranges; indeed, a slight limb brightening can be appreciated for ranges beyond $15$~\AA . No change in $\overline{I\,}$ is found when considering integration limits beyond $35$~\AA . 
%
 
The $\overline{Q}/\overline{I\,}$ signal remains relatively close to zero when considering only the contribution from the core region, as could be expected from the comparatively small linear polarization amplitude within this spectral range. The contribution from the interval between $200$ and $800$~m\AA , which contains the near-wing peaks, yields a very substantial increase to the $\overline{Q}/\overline{I\,}$ amplitude for all \gls*{los}s. 
More modest, but still appreciable,
increases in its amplitude are found when including the intervals between $0.8$ and $5$~\AA\ and between $5$ and $15$~\AA . Increasing the integration limit beyond $15$~\AA\ does not lead to any appreciable change in $\overline{Q}/\overline{I\,}$.  
These results illustrate that the $\overline{Q}/\overline{I\,}$ signal is dominated by the contribution from the negative peaks outside the line core, with a significant contribution coming from farther into the wings. 
As noted above, the magnetic sensitivity in such spectral regions is mainly due to \gls*{mo} effects, which therefore dominate the overall magnetic sensitivity of $\overline{Q}/\overline{I\,}$. 
Finally, bearing in mind that most of the near-wing photons come from the upper chromosphere and most of the photons farther into the wings originate from the middle chromosphere \citep[see Fig.~1 of][]{Belluzzi+12}, we conclude that the information encoded in $\overline{Q}/\overline{I\,}$ is mainly chromospheric. 
To ensure that we are accounting for all relevant wavelength contributions, we take integration limits of $50$~\AA\ from $\lambda_0$ for the broadband synthetic $\overline{I\,}$ and $\overline{Q}/\overline{I\,}$ signals in the rest of this paper.
 \subsection{Mimicking the CLASP broadband images with synthetic signals} 
 \label{subsec::DataSets}
In this subsection, we use the wavelength-integrated synthetic signals discussed above to mimic the large-\gls*{fov} broadband images obtained by CLASP (see Fig.~\ref{fig::CLASP_SJImage}). 
The considered \gls*{rt} code (see ABT22) is intended for \gls*{1d} plane-parallel static model atmospheres and therefore it cannot be used to reproduce the entire observed \gls*{fov} at once. Instead, each pixel of the \gls*{fov} is modeled separately (i.e., column by column). This \gls*{1d} approach cannot account for the impact of horizontal inhomogeneities in the atmospheric properties or bulk velocities with a horizontal component. As a further simplification, the pixels pertaining to the solar disk are categorized into four groups that correspond to regions of the solar atmosphere with different levels of activity, discriminating between them according to their intensity and distance from disk center (details can be found in Appendix~\ref{secApp::Construct}). 
We distinguish between internetwork regions (group \textit{i}), weakly (group \textit{ii}) and strongly (group \textit{iii}) magnetized network regions, and a plage region (group \textit{iv}). 
 
For the various groups, we used different atmospheric models according to the activity of the region we aim to replicate; we considered FAL-C for groups (\textit{i}) and (\textit{ii}), FAL-F for group (\textit{iii}), and FAL-P for group (\textit{iv}). 
We considered the Stokes profiles obtained from \gls*{rt} calculations both in the absence of magnetic fields and in the presence of fields whose strengths and orientations are representative of each group. For simplicity, such strengths and orientations were taken to be constant with height. 
For group (\textit{i}), an average was performed over a set of $16$ different profiles, each obtained from an \gls*{rt} calculation considering a $10$~G horizontal magnetic field, but with different, equally spaced, azimuths.\footnote{Performing such an average is not equivalent to taking a micro-structured magnetic field. Instead, it corresponds to the case in which the orientation of the magnetic field changes at scales above the mean free path of the line's photons but below the considered resolution element.}  
The resulting field distribution has no net longitudinal component, regardless of the considered LOS. 
A similar average over $16$ profiles from different magnetic field realizations was also performed for group (\textit{ii}), but considering $50$~G fields with a $30^\circ$ inclination with respect to the local vertical, which have a net longitudinal component for \gls*{los}s other than the vertical or completely horizontal directions. No such averages were performed for groups (\textit{iii}) and (\textit{iv}); in such cases the magnetic fields were taken to be vertical, with strengths of $125$ and $300$~G, respectively. 
The latter fields strengths are roughly $2.5$ and $6$ times the Hanle critical field, respectively. 
In the magnetized case, we considered two distinct scenarios for the pixels in group (\textit{ii}), in which the considered magnetic fields have a either positive or a negative projection along the local vertical (i.e., polarity). 
Similarly, for the pixels of group (\textit{iii}), we considered vertical magnetic fields with either positive or negative polarity. 
For group (\textit{iv}), we only considered fields with positive polarity. The polarity is of course undefined for group (\textit{i}), because in this case the magnetic field is horizontal. Each of the profiles was integrated over wavelength as discussed in the previous subsection. 

\begin{figure*}[!t]
 \centering 
\includegraphics[width=18cm]{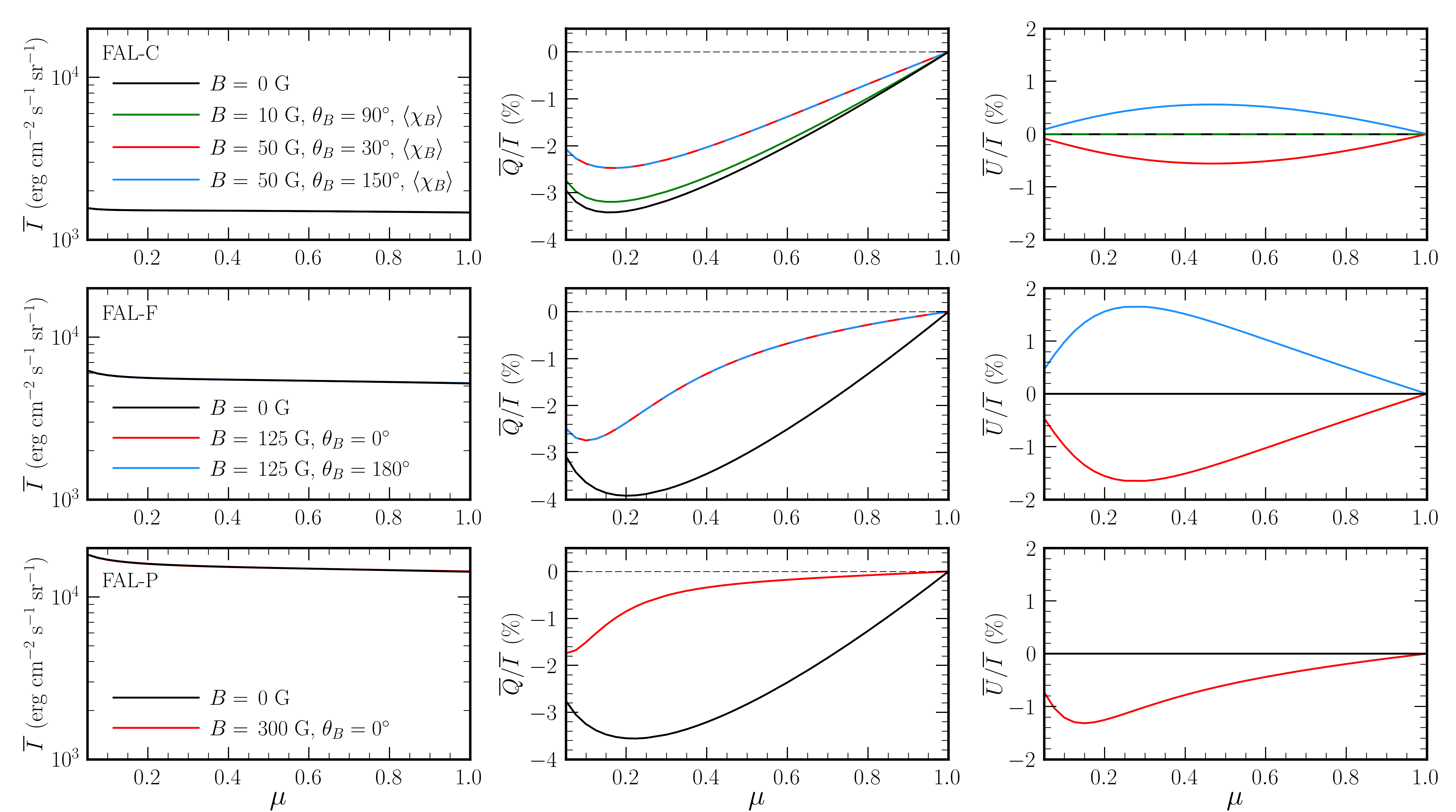}
 \caption{Broadband $\overline{I\,}$ (\textit{left column}), $\overline{Q}/\overline{I\,}$ (\textit{center column}), and $\overline{U}/\overline{I\,}$ (\textit{right column}) signals, obtained as explained in the text, as a function of the $\mu$ value of the LOS. The signals correspond to calculations for the FAL-C (groups \textit{i} and \textit{ii} discussed in the text; \textit{top row}), FAL-F (group \textit{iii}; \textit{middle row}), and FAL-P (group \textit{iv}; \textit{bottom row}) atmospheric models. The black curves represent the nonmagnetic case and the various colored curves (see legend) represent the magnetic fields considered for each group as discussed in the text. We recall that, for groups (\textit{ii}) and (\textit{iii}), fields with both positive (red curves) and negative (blue curves) polarities (see text) may be considered. The thin dashed black line in the center column indicates the zero value for $\overline{Q}/\overline{I\,}$. The reference direction for positive Stokes $Q$ is taken parallel to the limb.} 
 \label{fig::CLVfields}
 \end{figure*}

Figure \ref{fig::CLVfields} shows the \gls*{clv} for the $\overline{I\,}$, $\overline{Q}/\overline{I\,}$, and $\overline{U}/\overline{I\,}$ signals obtained from the above-mentioned profiles. The top row shows the \gls*{clv} for groups (\textit{i}) and (\textit{ii}), whereas the middle row shows the same for group (\textit{iii}), and the bottom row corresponds to group (\textit{iv}). The colored curves show the results of the calculations in which the magnetic fields considered for each group were included (showing both the positive and negative polarity case when applicable), whereas the black curves show the results for the nonmagnetic case. As discussed above, the wavelength-integrated intensity is not appreciably impacted by the magnetic field and, in practice, it depends only on the selected atmospheric model. 

\begin{figure*}[!t]
 \centering 
\includegraphics[width=18.0cm]{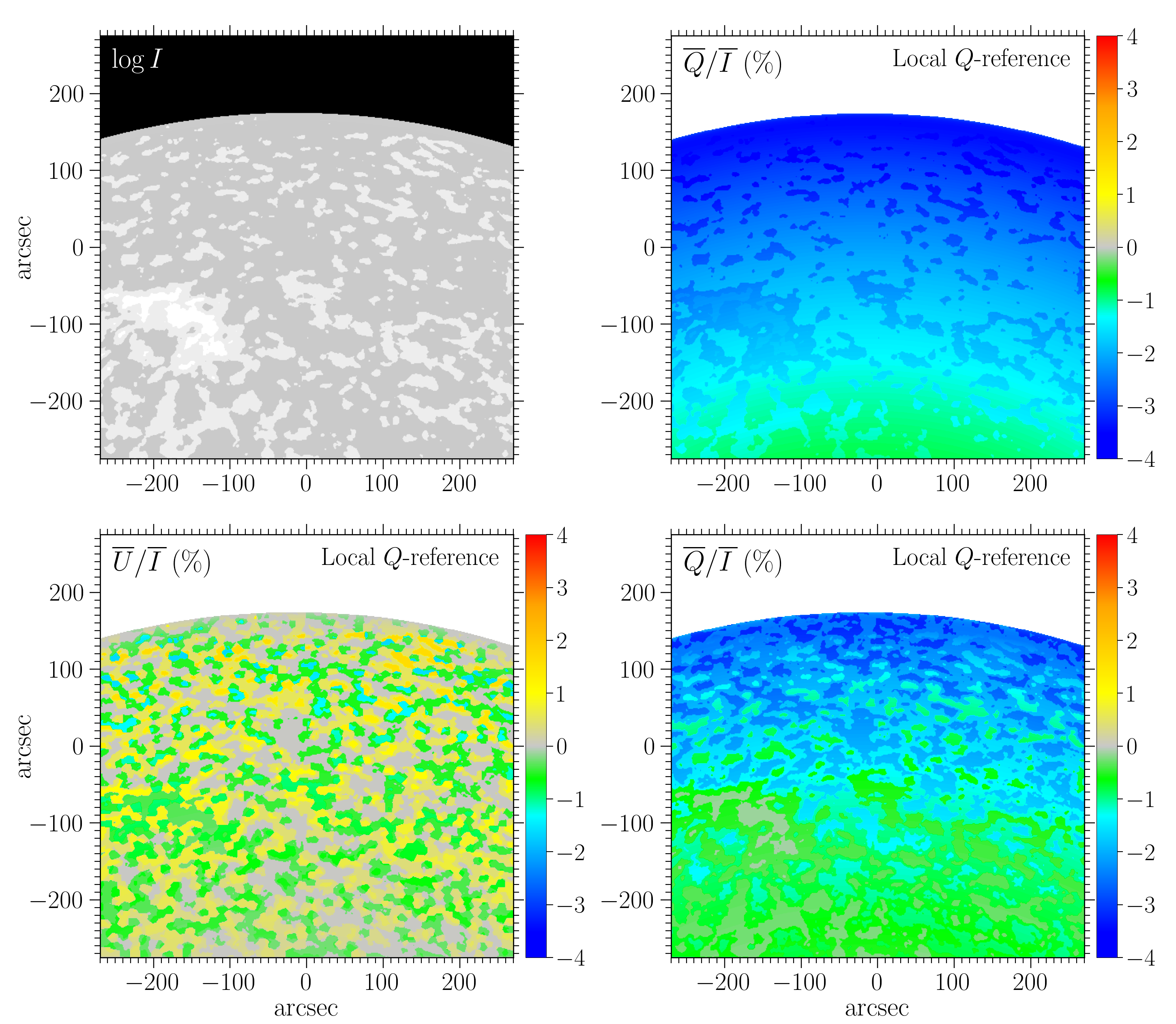}
 \caption{\textit{Upper panels}: $\overline{I \,}$ (\textit{left}) and ${\overline{Q}/\overline{I\,}}$ (\textit{right}) images obtained from the wavelength-integrated synthetic profiles obtained from \gls*{rt} calculations in the absence of magnetic fields. The images are constructed as described in the text and correspond to the region of the solar disk observed by the CLASP SJ. \textit{Lower panels}: ${\overline{U}/\overline{I\,}}$ (\textit{left}) and ${\overline{Q}/\overline{I\,}}$ (\textit{right}) images constructed from the wavelength-integrated profiles obtained from \gls*{rt} calculations accounting for the magnetic fields discussed in the text. The reference direction for positive Stokes $Q$ is taken to be perpendicular to the nearest limb (i.e., local $Q$-reference).} 
 \label{fig::SynthIQ}
\end{figure*} 
By contrast, the $\overline{Q}/\overline{I\,}$ signals are magnetically sensitive via the \gls*{mo} effects, as discussed above, and we verified that the same applies to the 
$\overline{U}/\overline{I\,}$ signals. 
For group (\textit{i}), the considered $10$~G magnetic field distribution has a very limited impact on the $\overline{Q}/\overline{I\,}$ signal. 
We recall that this magnetic distribution is characterized by a zero net longitudinal component for any \gls*{los}. Thus, the Stokes $U$ signals introduced by \gls*{mo} effects  perfectly cancel out upon performing the above-mentioned average over the profiles obtained from the various magnetic realizations. 
When considering instead the signals obtained for the slightly inclined $50$~G fields of group (\textit{ii}), $\overline{Q}/\overline{I\,}$ is depolarized to a greater, although still modest, extent. In this case, a considerable $\overline{U}/\overline{I\,}$ signal is found, and its sign depends on the polarity of the considered field distribution. A qualitatively similar behavior is found for the $125$~G vertical fields of group (\textit{iii}), although in this case the $\overline{Q}/\overline{I\,}$ signal is further depolarized and the amplitude of the $\overline{U}/\overline{I\,}$ signal is greater, especially for small $\mu$ values. Finally, when including the $300$~G positive-polarity vertical magnetic field of group (\textit{iv}), the amplitude of the $\overline{Q}/\overline{I\,}$ signal is greatly reduced. 
Moreover, the $\overline{U}/\overline{I\,}$ signal is smaller than the ones obtained for groups (\textit{ii}) and (\textit{iii}) in the magnetized case, as a consequence of the \gls*{mo}-effect-induced depolarization \citep[for a detailed discussion on such depolarization, see Appendix A of][]{AlsinaBallester+18}. 

Using the synthetic wavelength-integrated signals discussed above, we aimed to mimic the CLASP broadband images, taking the following steps. 
First, for each pixel of the observed image, we calculated the distance from disk center\footnote{Following \citet{Kubo+16}, the disk center is set at $(x_0,y_0) = (-17 \arcsec, -786\arcsec\!.5 )$.} and the associated $\mu$ value. The latter was determined according to 
\begin{equation}
\mu = \sqrt{1 - (r/R)^2} \, ,
\label{Eq:MuVal}
\end{equation}
where $r$ is the distance from the considered pixel to the solar disk center and $R = 961\arcsec\!.5 $ is the radius of the solar disk. 
Then, each pixel in the disk was assigned to one of the four groups according to its observed intensity value (see Appendix~\ref{secApp::Construct} for details on how this assignment was performed). 
As an additional condition to assign pixels to group (\textit{iv}), we required them to be in a rectangular area in the lower left quadrant of the image, where the plage is found.   
As noted above, only positive polarities were considered for the calculations of group (\textit{$iv$}), whereas both positive and negative polarities were considered for groups (\textit{$ii$}) and (\textit{$iii$}). 
As discussed in detail in Appendix~\ref{secApp::Construct}, once each pixel was assigned to one of the four groups, the polarities of all group--(\textit{ii}) and --(\textit{iii}) pixels were determined. First, these pixels were assembled into clusters according to their spatial position using a $k$-means algorithm. Then, all the pixels in a given cluster were assigned the same polarity (positive or negative), selected at random. For the figures shown in the rest of this subsection, $1400$ clusters were taken for the polarity selection. 
 
The filter-polarimetric images observed by CLASP were then mimicked by plotting the synthetic broadband signals according to the group, polarity, and $\mu$-value assigned to each pixel. As noted above, for the synthetic $Q/I$ and $U/I$ profiles, we take the reference direction for positive Stokes $Q$ parallel to the nearest limb by default. We note that, in observations over a large \gls*{fov} like those shown in Fig.~\ref{fig::CLASP_SJImage}, the direction of the nearest limb (and thus the reference direction for positive Stokes $Q$) changes according to the position on the solar disk. Hereafter, we refer to this particular choice of reference direction as local $Q$-reference. 

The resulting intensity image, obtained without accounting for the magnetic field in the \gls*{rt} calculations, is shown in the upper left panel of Fig.~\ref{fig::SynthIQ}. No appreciable change was found in the intensity image when accounting for the magnetic fields, and therefore the corresponding image is not shown in the figure. 
The $\overline{Q}/\overline{I\,}$ images in the local $Q$-reference, obtained in the absence and in the presence of magnetic fields, are shown in the upper and lower right panels, respectively. 
In both cases, we observe an increase in the amplitude as the distance from the limb (or $\mu$ value) decreases, in agreement with what is shown in Fig.~\ref{fig::CLVfields}. 
The impact of the magnetic field through the \gls*{mo} effects is apparent when comparing the two images. In the absence of magnetic fields, the polarization amplitudes of the plage and network regions are slightly larger than those of the internetwork regions at the same $\mu$. The smallest polarization amplitudes, of about $0.6\%$, are found in the internetwork regions closest to the disk center, whereas the amplitudes in the plage region exceed $1\%$. On the other hand, when the magnetic fields are included, the signals are
strongly depolarized in the plage and strongly magnetized network regions, so that their amplitudes become considerably smaller than in the surrounding regions. A substantial  depolarization is also found in the weakly magnetized network regions, and a slight depolarization is even appreciable in the internetwork region. This leads to a far better qualitative and quantitative agreement with the image observed by CLASP, which suggests that the depolarized features, found in the regions of the solar disk where higher levels of magnetic activity are expected, are compatible with the action of \gls*{mo} effects. 

These results, however, do not exclude that other physical mechanisms can play a critical role in the observed polarization signals. Indeed, the observed broadband $Q/I$ signals are found to have, on average, lower amplitudes than the image constructed from synthetic wavelength integrated images, even when accounting for the impact of the magnetic fields, both at limb positions and closer to disk center (contrast Fig.~\ref{fig::CLASP_SJImage} and Fig.~\ref{fig::SynthIQ}). This could be a consequence of the fact that the symmetry breaking effects of 3D RT, which are known to impact the linear polarization produced by scattering processes, are being neglected. 

In the absence of magnetic fields, the axial symmetry of the problem is preserved in the considered \gls*{1d} \gls*{rt} modeling and thus $\overline{U}/\overline{I\,}$ is zero throughout the \gls*{fov} when considering the local $Q$-reference. When magnetic fields are included, nonzero $\overline{U}/\overline{I\,}$ signals are produced by \gls*{mo} effects, as shown in the lower left panel of Fig.~\ref{fig::SynthIQ}. Similar to what occurs for $\overline{Q}/\overline{I\,}$, the amplitude of $\overline{U}/\overline{I\,}$ tends to increase when approaching the limb, where larger scattering polarization signals are produced (see Fig.~\ref{fig::CLVfields}). 
In the internetwork regions, the $\overline{U}/\overline{I\,}$ signal is zero, because the magnetic field distribution considered for group (\textit{i}) has no net longitudinal component. On the other hand, $\overline{U}/\overline{I\,}$ signals with a modest amplitude are found in the group-(\textit{ii}) network regions, which have relatively weak fields. 
Larger amplitudes are found in the strongly magnetized network and plage regions, although they are smaller in the latter regions because of the \gls*{mo}-effect-induced depolarization. 
Of course, the sign of the $\overline{U}/\overline{I\,}$ signals depends on the magnetic field polarity at the considered pixel. 
The relatively sharp edges that separate regions of positive and negative $\overline{U}/\overline{I\,}$ are signatures of the $k$-means algorithm used to assign different polarities to clusters of pixels (see discussion above in this section; further details are given in Appendix~\ref{secApp::Construct}). 

 \begin{figure}[!h]
 \centering 
\includegraphics[width=8.25cm]{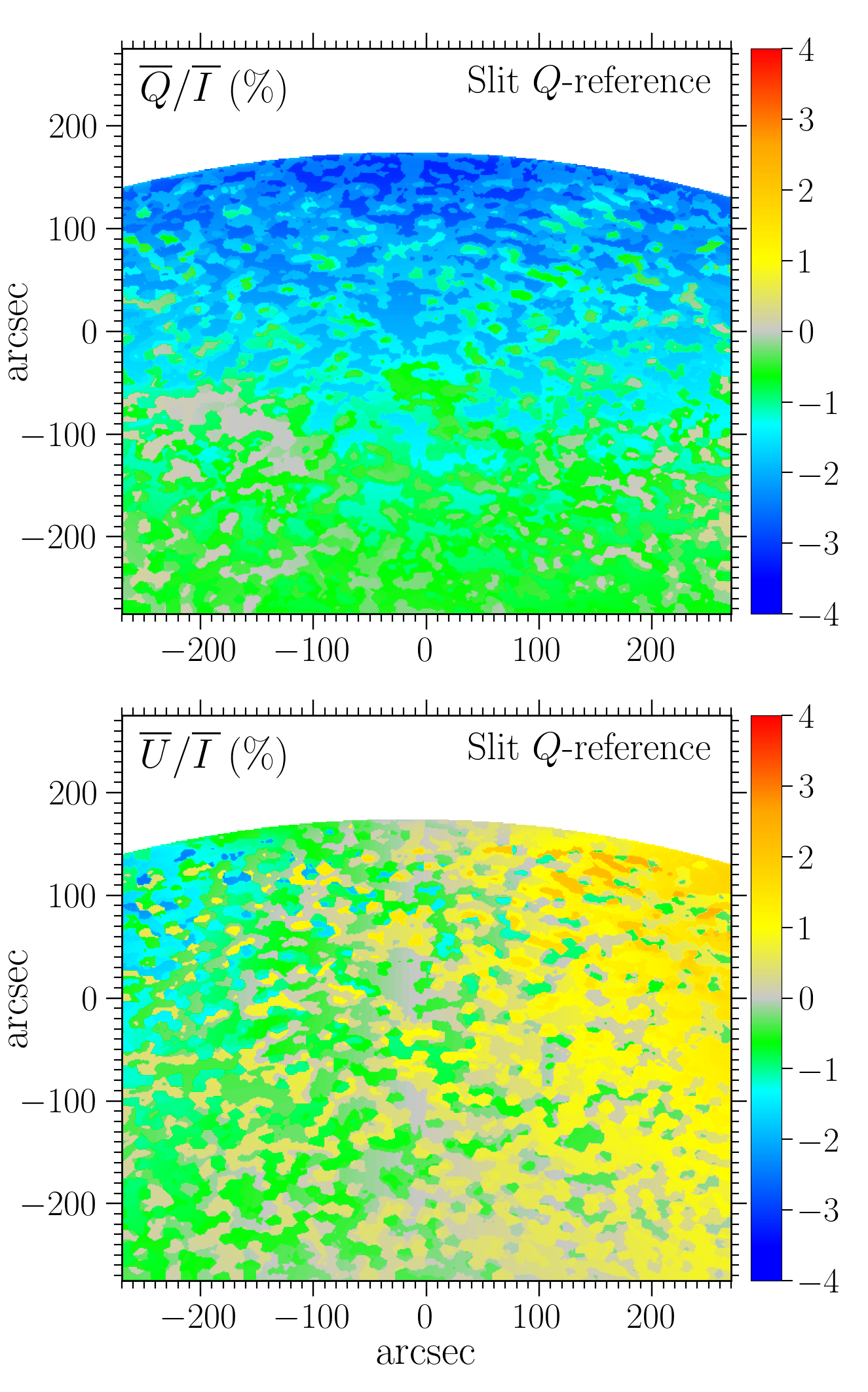}
 \caption{Synthetic $\overline{Q}/\overline{I\,}$ \textit{(upper panel)} and $\overline{U}/\overline{I\,}$ \textit{(lower panel)} images constructed as discussed in the text, accounting for the magnetic fields given in Sect.~\ref{subsec::DataSets} in the \gls*{rt} calculations, and considering the reference direction for Stokes $Q$ parallel to the slit (i.e., slit $Q$-reference).} 
 \label{fig::Synth_QU}
 \end{figure}

Typically, polarimetric observations are carried out considering the same reference direction for positive Stokes $Q$ at all points in the image. Indeed, for the CLASP broadband observations shown in Fig.~\ref{fig::CLASP_SJImage}, the reference direction for positive Stokes $Q$ was perpendicular to the slit (i.e., slit $Q$-reference) throughout the \gls*{fov}. Transforming the image so that the reference direction for positive $Q$ varies with the spatial position requires considering different combinations of the (slit $Q$-reference) Stokes $Q$ and $U$ signals at different points. The transformation of the images from the local $Q$-reference to the slit $Q$-reference and vice versa is detailed in Appendix~\ref{secApp::Qref}. 

The $\overline{Q}/\overline{I\,}$ and $\overline{U}/\overline{I\,}$ images, constructed from synthetic profiles in which the magnetic fields are taken into account as explained above 
and transformed into the slit $Q$-reference, are shown in Fig.~\ref{fig::Synth_QU} (compare with the same images in the local $Q$-reference given in the lower panels of Fig.~\ref{fig::SynthIQ}). 
The most apparent differences between the linear polarization images in the local and slit $Q$-reference are found at the positions where the angle $\alpha$ between the slit and the local radial direction is the largest, namely at the lower right and left edges of the \gls*{fov}.
The expression for $\alpha$ can be found in Appendix~\ref{secApp::Qref}. 
In the slit $Q$-reference, a preference for positive (negative) $\overline{U}/\overline{I\,}$ signals is found on the right (left) side of the image, in accordance with the sign of $\alpha$, because of the contribution from the $\overline{Q}/\overline{I\,}$ in the local $Q$-reference, which is negative throughout the \gls*{fov}. 
We also verified that, in the absence of magnetic fields, a nonzero $\overline{U}/\overline{I\,}$ image is found in the slit $Q$-reference; in this case, its sign depends only on that of $\alpha$. 
Likewise, the amplitude of $\overline{Q}/\overline{I\,}$ in the slit $Q$-reference increases or decreases, relative to that in the local $Q$-reference, at each point in the \gls*{fov} depending on the sign of the product of $\alpha$ and $\overline{U}/\overline{I\,}$ in the local $Q$-reference. 
Again, the relatively sharp edges found between different clusters of pixels in the $\overline{U}/\overline{I\,}$ image in Fig.~\ref{fig::Synth_QU} -- and to a smaller extent in the  $\overline{Q}/\overline{I\,}$ image -- are signatures of the $k$-means algorithm that was used used to select the polarity of group (\textit{ii}) and (\textit{iii}) pixels. 
The dependency of $\overline{Q}/\overline{I\,}$ and $\overline{U}/\overline{I\,}$ on the total number of clusters is illustrated in Fig.~\ref{fig::AppQUClust} in Appendix~\ref{secApp::Construct}. 

We recommend that the researcher interested in studying solar longitudinal magnetic fields 
through filter-polarimetric observations over a large \gls*{fov} first transforms the image into a local $Q$-reference, for instance parallel to the nearest limb (as in this work) or perpendicular to the nearest limb. 
Considering such $Q$-references, the degree of rotation of the plane of linear polarization given by $\overline{Q}/\overline{I\,}$ and $\overline{U}/\overline{I\,}$ at each spatial point in the image can be more easily quantified. 

Finally, the polarization fraction $P_L = \sqrt{\overline{Q}^2 + \overline{U}^2}/\overline{I\,}$ can provide a complementary view of the variation of the linear polarization signal throughout the \gls*{fov} and could be used to study its variation with respect to a given reference value. 
Figure~\ref{fig::Synth_PL} shows the synthetic $P_L$ image obtained when accounting for the magnetic field (i.e., corresponding to the lower panels of Fig.~\ref{fig::SynthIQ} and to Fig.~\ref{fig::Synth_QU}).  
As expected, the $P_L$ amplitude tends to increase toward the limb, and it locally decreases in regions with larger magnetic fields due to the depolarizing action of \gls*{mo} effects. 
The smallest amplitudes are thus found in the plage area, where a vertical field of $300$~G was considered. 
The $P_L$ amplitude in a filter-polarimetric Ly$\alpha$ image, together with degree of rotation of the plane of linear polarization, encodes highly valuable information on the thermodynamical properties in the mid-upper layers of the solar chromosphere. 

\begin{figure}[!h]
 \centering
\includegraphics[width=8.25cm]{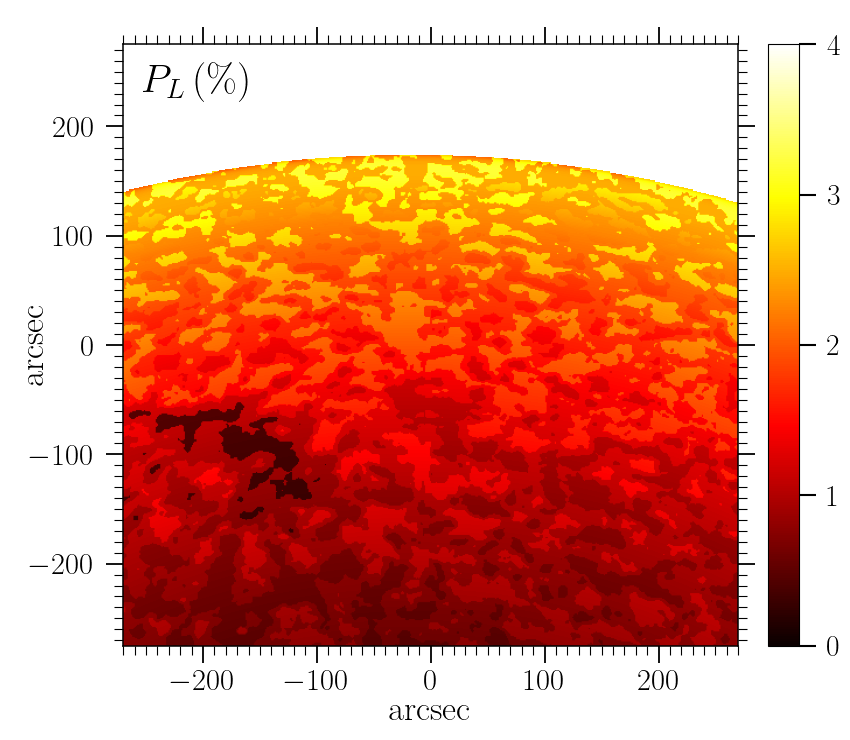}
\caption{Linear polarization fraction $P_L$ image constructed as discussed in the text, accounting for the magnetic fields given in Sect.~\ref{subsec::DataSets} in the \gls*{rt} calculations.}
\label{fig::Synth_PL}
\end{figure} 

\section{Conclusions}
\label{sec::Conclusions}
We investigated the potential of the 
broadband intensity and linear polarization signals, centered on the hydrogen Ly$\alpha$ line, for extracting information on the thermodynamic and magnetic structure of the solar chromosphere. 
We applied a non-\gls*{lte} \gls*{rt} code that we recently developed, which can suitably synthesize the intensity and polarization profiles of the Ly$\alpha$ line, considering a two-term atomic model and accounting for \gls*{prd} effects, $J$-state interference, and magnetic fields of arbitrary strength and orientation (for details, see ABT22). 
We considered synthetic profiles corresponding to regions of the solar atmosphere with different levels of activity, obtained from different \gls*{1d} semi-empirical atmospheric models. We integrated such profiles over wavelength, weighting them with a Gaussian with a \gls*{fwhm} of $35$~\AA , as in the broadband images taken by the CLASP mission \citep{Kano+17}. 
The most important contribution to the wavelength-integrated linear polarization signals was found to come from spectral range between $0.2$ and $0.8$~\AA , whose radiation comes mainly from the upper chromosphere, and the remaining contribution comes mostly from further into the wings, whose radiation comes mainly from the middle chromosphere. Thus, these magnetically sensitive signals primarily encode information on the middle-upper layers of the solar chromosphere. 

For illustrative purposes, we used wavelength-integrated synthetic profiles to mimic the broadband images taken by the CLASP mission over a large \gls*{fov}. In each pixel of the generated image, we used the synthetic data obtained from \gls*{rt} calculations  considering atmospheric models and magnetic fields that were selected according to the intensity value and limb distance of the corresponding pixel in the observed image.  
The calculations were carried out considering one independent column for each pixel, without accounting for horizontal transfer of radiation between them.  
For brighter regions of the solar disk, associated with a higher degree of activity, the corresponding profiles were obtained considering the FAL-P or FAL-F atmospheric models, rather than FAL-C. 
For the network and plage regions, vertical magnetic fields of greater strength were considered. 
The inclusion of magnetic fields allows for a much better agreement between the synthetic and observed filter-polarimetric images. 
In addition to the increase in amplitude as the solar limb is approached (found also in the nonmagnetic case), the synthetic $\overline{Q}/\overline{I\,}$ signals are depolarized by \gls*{mo} effects in the regions of higher activity, especially those corresponding to the plage area. The fact that only a qualitative agreement with observations could be achieved was expected, because the synthetic images were constructed through a simplified approach that considers only a small selection atmospheric models and magnetic fields. It is also worth noting that the horizontal variations in the thermodynamical properties of the solar atmosphere were ignored for simplicity, although they are expected to modify the polarization signals by introducing an axial asymmetry. 
Regardless, the qualitative agreement with the two-dimensional broadband image observed by CLASP suggests 
that certain large-scale features such as the depolarization found in plage and network regions can be explained, at least in part, by the action of \gls*{mo} effects. 

The \gls*{mo} effects have a clear impact not only on the broadband $Q/I$ signals, but also on the $U/I$ signals, because they produce a rotation of the plane of linear polarization. Although the CLASP \gls*{sj} system did not provide broadband $U/I$ images, the spectropolarimeter revealed nonzero $U/I$} signals in the wings as well as in the core \citep[see Figs.~1 and 3 of][]{Kano+17}. 
Such observations revealed spatial variations and even sign changes along the slit, in apparent agreement with the large-\gls*{fov} synthetic $\overline{U}/\overline{I\,}$ images presented in this work. 

Clearly, the broadband linear polarization signals centered on the Ly$\alpha$ line have considerable diagnostic potential for probing the thermodynamical and magnetic properties of the middle-upper layers of the solar chromosphere over a large \gls*{fov}. However, in order to reliably infer quantitative information about longitudinal chromospheric magnetic fields (via \gls*{mo} effects) and other properties, it is essential to account for the full three-dimensional complexity of the solar atmosphere, including the horizontal variations in properties such as temperature and density. The development of a practical non-\gls*{lte} code that can account for \gls*{3d} \gls*{rt}, as well as the \gls*{prd} effects and $J$-state interference required to suitably model the wing linear polarization of many strong resonance lines, is a challenge, but in forthcoming publications we will stepwise address this important question.  \\

We acknowledge the funding received from the European Research Council (ERC) under the European Union's Horizon 2020 research and innovation program (ERC Advanced Grant agreement No. 742265). E.A.B. and L.B. gratefully acknowledge financial support by the Swiss National Science Foundation (SNSF) through Grant 200021\_175997.  
L.B. and J.T.B. acknowledge financial support through the Sinergia program of the SNSF (Grant No. CRSII5\_180238).  

\appendix

\section{Linear polarization signals and reference direction for $Q > 0$}
\label{secApp::Qref}
\begin{figure*}[!h]
 \centering 
\includegraphics[width=12.0cm]{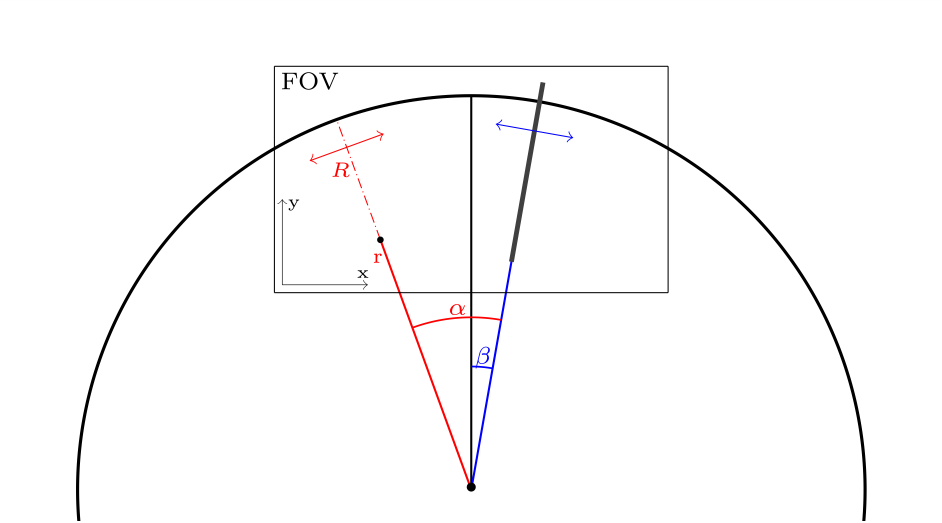}
 \caption{Schematic representation of the observation of the solar disk taken by CLASP and shown in Fig.~\ref{fig::CLASP_SJImage}. 
 The box represents the \gls*{fov} of the \gls*{sj} system presented in that figure. The black line, which goes from the disk center to the limb, is parallel to the $y$-axis. 
 The blue line is parallel to the slit, which is represented by the thick dark gray line. The radial direction is given by the disk center and the spatial point under consideration (marked by black dots), which are connected by the solid red line of length $r$. The dashed-dotted red line is parallel to the radial direction and has the length of the disk radius $R$. $\alpha$ is the angle between the slit and the radial direction. $\beta$ is the angle between the slit and the $y$-axis. The blue two-headed arrow is perpendicular to the slit. The red two-headed arrow is perpendicular to radial direction or, equivalently, parallel to the limb. } 
 \label{fig::Diagram}
 \end{figure*}

Two different conventions for the definition of Stokes $Q$ and $U$ are used in the various two-dimensional synthetic images presented in Sect.~\ref{subsec::DataSets}. In the first convention (i.e, the local $Q$-reference), the direction for positive Stokes $Q$ is parallel to the nearest limb or, equivalently, perpendicular to the line between the considered point and the disk center. Clearly, the direction varies from point to point in the \gls*{fov}. In the second convention (i.e., the slit $Q$-reference), the direction for positive $Q$ is perpendicular to the slit direction, and is therefore the same at all points in the \gls*{fov}. At a given spatial point, the $Q$ and $U$ Stokes parameters in the local $Q$-reference can be transformed to those in the slit $Q$-reference through the rotation  
\begin{subequations}
 \begin{align}
   Q_{\mbox{\scriptsize slit}} & = Q_{\mbox{\scriptsize local}} \cos{2\alpha} + U_{\mbox{\scriptsize local}} \sin{2\alpha} \, , \\
   U_{\mbox{\scriptsize slit}} & = - Q_{\mbox{\scriptsize local}} \sin{2\alpha} + U_{\mbox{\scriptsize local}}  \cos{2\alpha} \, ,
   \end{align}
  \label{ApEqSlitTransf}
\end{subequations}
where $\alpha$ is the angle between the slit direction and the radial direction (i.e., the line between the disk center and the spatial point under consideration) or, equivalently, the angle between the direction perpendicular to the slit and direction tangent to the nearest limb. 
The inverse transformation, from the slit $Q$-reference to the local $Q$-reference, corresponds to the rotation in the opposite direction. As such, it is trivial to obtain it from the previous expression (simply by changing the sign of $\alpha$) and it is not shown here. 
Taking $\beta$ to be the angle between the slit and the $y$-axis in the observed two-dimensional image of CLASP (see Fig.~\ref{fig::CLASP_SJImage}), 
\begin{equation}
 \alpha = \arcsin \bigl( \Delta x/r \bigr) - \beta \, , 
\end{equation}
where $r$ is the distance between the point under consideration and the disk center, which have coordinates $(x,y)$ and $(x_0,y_0)$, respectively, and where $\Delta x = x - x_0$. 
For the considered CLASP image, $(x_0,y_0) = (-17 \arcsec, -786.5 \arcsec)$ and $\beta \simeq 0.5^\circ$. 
A schematic representation of the angles and directions defined here is given in Fig.~\ref{fig::Diagram}. 

\section{Mimicking filter-polarimetric images with synthetic data}
\label{secApp::Construct} 
As discussed in Sect.~\ref{subsec::DataSets}, we prepared the synthetic broadband images by assigning to each on-disk pixel its corresponding $\mu$ value (given by Eq.~\eqref{Eq:MuVal}), one of the groups introduced in Sect.~\ref{subsec::DataSets}, and (in the case of pixels in groups \textit{ii} and \textit{iii}) the polarity of the magnetic field. 
Subsequently, the Stokes images were prepared by filling in each pixel with the wavelength-integrated, $\mu$-dependent, Stokes signals obtained from \gls*{rt} calculations, carried out considering the atmospheric model and magnetic field that correspond to the group and polarity assigned to the pixel. 
The wavelength integration limits were taken to extend $50$~\AA\ from the line center and a Gaussian weighting function with a \gls*{fwhm} of $35$~\AA\ was employed. In this Appendix, we provide further details on how the group and the magnetic field polarity (when applicable) were assigned to each pixel. 

\begin{figure}[!h]
 \centering
\includegraphics[width=8.25cm]{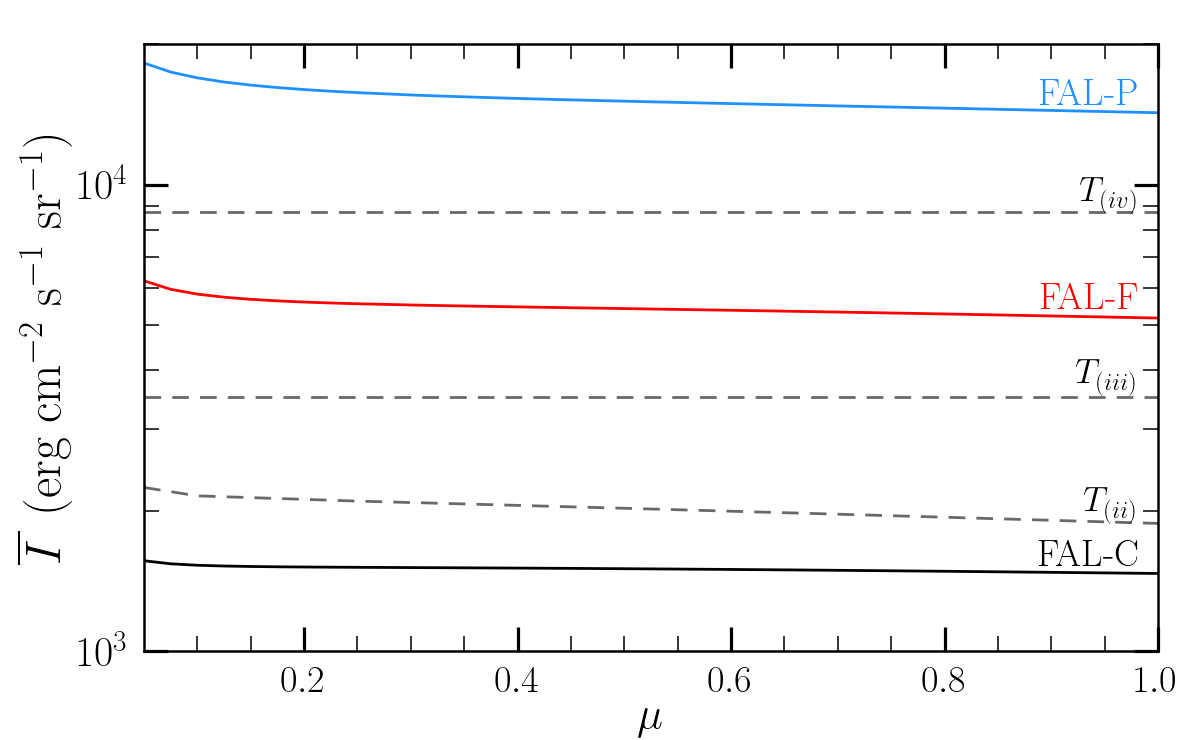}
\caption{Colored curves: CLV for the synthetic $\overline{I\,}$ obtained from RT calculations considering the FAL-C, FAL-F, and FAL-P atmospheric models (see labels), also shown in Fig.~\ref{fig::CLVfields}. Grey dashed curves: Intensity thresholds for assigning a given pixel of the image to groups (\textit{ii}), (\textit{iii}), and (\textit{iv}), as indicated by the corresponding label. Such thresholds are linear piecewise functions of $\mu$ whose coefficients are given in Table.~\ref{tab:Table1}. 
In order to directly confront the $T_X(\mu)$ thresholds shown in this figure with the aforementioned synthetic intensities, we substitute the $\overline{I\,}_{\!\mbox{\tiny max}}$ introduced in the text with the synthetic value of $\overline{I\,}$ obtained with FAL-P at $\mu = 0.05$  in the $c_X^Y$ coefficients.} 
\label{fig::FigApp1}
\end{figure} 

All pixels are set to group (\textit{i}) by default, and are only assigned to groups (\textit{ii}), (\textit{iii}), and finally (\textit{iv}) if the intensity of the associated pixel in the CLASP SJ image (see upper panel of Fig.~\ref{fig::CLASP_SJImage}) exceeds the value of three increasingly restrictive thresholds $T_X$, with $X = \bigl\{(ii), (iii), (iv) \bigr\}$, respectively. As noted below and in the main text, pixels could only be assigned to group (\textit{iv}) if they additionally met the requirement that they are inside a rectangular area in the lower left quadrant of the image. 
As noted in the main text, the thresholds are given as piecewise linear functions of $\mu$, 
\begin{align}
  T_X(\mu) & = 
    \begin{dcases}
     c_X^a + (c_X^b - c_X^a) \, \frac{\mu - 0.05}{0.05} \,,& \mu < 0.1 \, ; \\
     c_X^b + (c_X^c - c_X^b) \, \frac{\mu - 0.1}{0.25} \,,& 0.1 \leq \mu < 0.25 \, ; \\
     c_X^c + (c_X^d - c_X^c) \, \frac{\mu - 0.25}{0.75} \,,&  0.25 \leq \mu \, .
    \end{dcases}
\end{align}
The $c_X^Y$ coefficients were selected to yield the best qualitative agreement between the synthetic intensity image in the left panel of Fig.~\ref{fig::SynthIQ} and the observed intensity image in the upper panel of Fig.~\ref{fig::CLASP_SJImage}. Their values are shown in Table~\ref{tab::TabCoef}, where they are normalized to the maximum intensity in the observed broadband image $\overline{I\,}_{\!\mbox{\tiny max}}$.  
\begin{table}[!h] 
 \centering 
 \caption{\label{tab:Table1} Threshold function coefficients \textcolor{black}{normalized to $\overline{I\,}_{\!\mbox{\tiny max}}$}} 
 \renewcommand{\arraystretch}{1.3}
 \begin{tabular}{|c|c c c c|} \hline 
 $c_X^Y/\overline{I\,}_{\!\mbox{\tiny max}}$ & $Y = \mathrm{a}$ & $Y = \mathrm{b}$ & $Y = \mathrm{c}$ & $Y = \mathrm{d}$ \\ \hline
 \renewcommand{\arraystretch}{1}
 $X = (ii)$  &  $0.123$ &  $0.118$ & $0.115$ & $0.103$ \\
 $X = (iii)$ &  $0.192$ &  $0.192$ & $0.192$ & $0.192$ \\
 $X = (iv)$  &  $0.480$ &  $0.480$ & $0.480$ & $0.480$  \\ \hline
 \end{tabular}
 \label{tab::TabCoef}
\end{table}

Figure \ref{fig::FigApp1} shows a comparison between the threshold intensities and the synthetic $\overline{I\,}$ signals resulting from the atmospheric models considered for the various groups, given as a function of $\mu$. 
In order facilitate a direct comparison with the synthetic signals, the $T_X(\mu)$ shown in the figure were obtained using $c_X^Y$ coefficients like those shown in Table~\ref{tab::TabCoef}, but substituting $\overline{I\,}_{\!\mbox{\tiny max}}$ with the synthetic $\overline{I\,}$ value obtained with FAL-P at $\mu = 0.05$. We recall that the data sets for groups (\textit{ii}), (\textit{iii}), and (\textit{iv}) were obtained considering three different FAL models (C, F, and P, respectively). As can be seen from the colored curves in Fig.~\ref{fig::FigApp1}, the wavelength-integrated intensity signals obtained from all three models present a very modest \gls*{clv}, which is only appreciable at small $\mu$ values. Thus, we find it a reasonable choice to take the thresholds $T_{(iii)}$ and $T_{(iv)}$, which distinguish between these groups, to be fully independent of $\mu$. 
We note that, when preparing the synthetic images shown in the main text, the threshold to assign pixels to group (\textit{iv}) was only applied to those in the rectangular region within the FOV such that $x \in (-270\arcsec, -55\arcsec)$ and $y \in (-176\arcsec, 0 \arcsec)$, which contains the entirety of the plage area. 
Unlike the other two thresholds, $T_{(ii)}$ distinguishes between groups whose associated data sets were obtained with the same atmospheric model (FAL-C) and which only differ in the choice of magnetic field (see Sect.~\ref{subsec::DataSets} for the details on the modeling of pixels in groups \textit{i} and \textit{ii}). 
For this threshold, the $c_{(ii)}$ coefficients were selected so that the piecewise function is not flat with $\mu$, but instead presents a clear increase 
as smaller $\mu$ values are approached. This made it possible to achieve a greater degree of local variation in the $\overline{Q}/\overline{I\,}$ image close to the limb, in better agreement with observations. 

\begin{figure*}[!t]
 \centering 
\includegraphics[width=18cm]{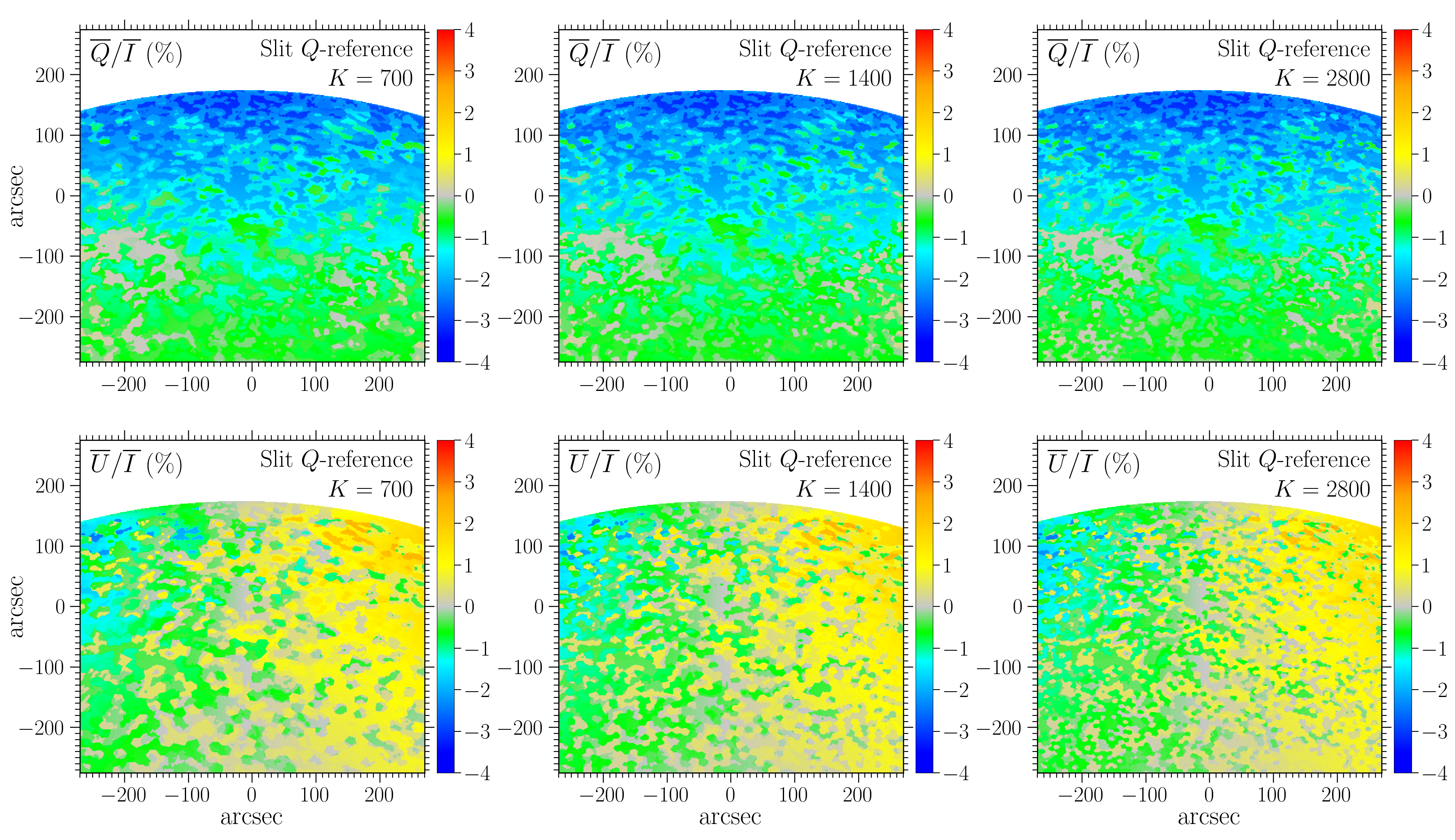}
 \caption{Synthetic $\overline{Q}/\overline{I\,}$ \textit{(upper panels)} and $\overline{U}/\overline{I\,}$ \textit{(lower panels)} images constructed as discussed in the text, accounting in the RT calculations for the magnetic fields given in Sect.~\ref{subsec::DataSets}, and considering the slit $Q$-reference. The images were constructed using a $k$-means algorithm to assign positive or negative polarities to clusters of pixels that correspond to groups (\textit{ii}) and (\textit{iii}) as discussed in the text, taking a total of $700$ clusters (\textit{left panels}), $1400$ as in the main text (\textit{center panels}), and $2800$ (\textit{right panels}).} 
 \label{fig::AppQUClust}
 \end{figure*} 
Magnetic fields in the solar atmosphere are often analyzed using magnetographs, which are related to measurements of the circular polarization. The sign of the circular polarization determines the polarity of the fields. Magnetograph observations reveal that structured magnetic fields of positive or negative polarity are found throughout the solar disk, even in quiet regions where narrow and elongated magnetic field concentrations are found at various spatial scales, up to those comparable to super-granulation \citep[see review by][and references therein]{BellotRubioOrozcoSuarez19}. Suitably reproducing such features and their structured nature requires a highly sophisticated magnetoconvective modeling of the solar atmosphere and the underlying convection zone, which is beyond the scope of the present work. Indeed, instead of selecting the polarity of every pixel in the image through a physically guided technique, we chose a relatively quick approach that yields a distribution of polarities that resembles what is expected from observations. We remind the reader that the goal of this paper is to highlight the diagnostic potential of the wavelength-integrated Ly$\alpha$ linear polarization signals, rather than to accurately reproduce observations. 

The assignment of a polarity to each pixel depends on the group into which it was sorted. No polarity was assigned to the pixels of group (\textit{i}) because, even when magnetic fields are included in the calculations, they are horizontal and thus their projection onto the local vertical is zero. On the other hand, we assigned a positive polarity to all pixels in group (\textit{iv}). For the pixels in the remaining groups (\textit{ii}) and (\textit{iii}) -- those corresponding to weakly and strongly magnetized network regions -- we enforced spatial structure in the assignment of their polarities via a $k$-means clustering. 
We used the routine provided by the \texttt{python} library \texttt{scikit-learn} \citep[see][]{scikref} which, in the default setting that we considered, makes use of Lloyd's algorithm \citep{Lloyd82}. 
Through this approach, the pixels that belong to the network groups were separated into clusters according to their $(x,y)$ position. In each of the $K$ considered clusters, the polarity was selected to be either positive or negative at random, and the same polarity was assigned to all the pixels in the selected cluster. 
We took the probability that the pixels of given cluster have negative polarity to be $p = 0.55$, making the assumption that the net flux over the \gls*{fov} should be roughly zero, and taking into account that we assigned a positive polarity to all the pixels of group (\textit{iv}). 
Prior to considering $k$-means clustering, we had attempted to assign the polarities using a flooding algorithm that ensured all adjacent group-(\textit{iii}) pixels have the same polarity and, subsequently, assigned the polarity of the group-(\textit{ii}) pixels in terms of the polarity of the nearest group-(\textit{iii}) and -(\textit{iv}) pixels, using a Gaussian blurring. However, the qualitative agreement between the synthetic and the observed $Q/I$ broadband images was significantly worse than the one reached when using the $k$-means approach. The synthetic images obtained with the flooding algorithm are thus not shown in this Appendix. 

The images shown in Sect.~\ref{subsec::DataSets} were constructed from wavelength-integrated synthetic signals as discussed above, assigning polarities to the pixels by setting the number of clusters to $K = 1400$ in the $k$-means routine. Figure~\ref{fig::AppQUClust} illustrates the impact of the choice of $K$ on the resulting broadband linear polarization images. The figure shows a comparison of the $\overline{Q}/\overline{I\,}$ and $\overline{U}/\overline{I\,}$ images, in the slit $Q$-reference, obtained when taking $K = 700$, $1400$, and $2800$. For all three cases, sharp boundaries are found between the clusters, which are especially appreciable in the images for $\overline{U}/\overline{I\,}$ and are clearly an artifact of the clustering algorithm. 
For the sake of simplicity, the polarity in each cluster was randomly assigned and was thus completely independent of that of the surrounding clusters, but relatively large concentrations of pixels with the same polarity, spanning multiple clusters, are still appreciable. This is somewhat obscured in the slit $Q$-reference, due to the mixing between the $\overline{Q}/\overline{I\,}$ and $\overline{U}/\overline{I\,}$ images in the local $Q$-reference and the preference for positive (negative) signals in the right (left) side in the latter image. However, such boundaries can be better appreciated in the $\overline{U}/\overline{I\,}$ images in the local $Q$-reference, which is shown in Fig.~\ref{fig::Synth_PL} for the $K = 1400$ case. 
This apparent spatial coherence is found to a greater extent when considering a smaller number of clusters $K$, each of which is of course larger in area. For the images present in the main text we selected $1400$ clusters, in order to reach the best balance between the local variations in linear polarization, as observed by CLASP (see Fig.~\ref{fig::CLASP_SJImage}), and the expected scale of single-polarity concentrations in the chromospheric network. 
\bibliographystyle{aasjournal}
\bibliography{cbib}
\end{document}